\documentclass[iop]{emulateapj}
\usepackage{graphicx}

\newcommand{\dqzero}{SDSS J1337--0026}
\newcommand{\dqone}{SDSS J2200--0741}
\newcommand{\dqtwo}{SDSS J2348--0942}
\newcommand{\dqtx}{SDSS J1426+5752}

\shorttitle{A New Small-Amplitude Variable Hot DQ White Dwarf}
\shortauthors{Dunlap, Barlow, Clemens}

\begin{document}

\title{A New Small-Amplitude Variable Hot DQ White Dwarf\footnotemark[*]}\footnotetext[*]{Based on observations at the SOAR Telescope, a collaboration between CPNq-Brazil, NOAO, UNC, and MSU.}

\author{B. H. Dunlap\footnotemark[$\dagger$], B. N. Barlow\footnotemark[$\dagger$], \& J. C. Clemens}\footnotetext[$\dagger$]{Visiting astronomer, Cerro Tololo Inter-American Observatory, National Optical Astronomy Observatory, which is operated by the Association of Universities for Research in Astronomy, under contract with the National Science Foundation.}
\affil{Department of Physics and Astronomy, University of North Carolina, Chapel Hill, NC 27599, USA\\
bhdunlap@physics.unc.edu}

\slugcomment{Accepted for publication in the Astrophysical Journal Letters}

\begin{abstract}
We present the discovery of photometric variations in the carbon-dominated atmosphere (hot DQ) white dwarf star SDSS J133710.19--002643.6.  We find evidence for two low-amplitude, harmonically-related periodicities at 169 s and 339 s, making it the fastest and smallest amplitude of the four known hot DQ variables and the only variable whose spectrum suggests the presence of hydrogen.  The star's fundamental and harmonic amplitudes are roughly equal, and its pulse shape is similar to the other two members of the class with detected harmonics.  Like the other variables, it appears relatively stable in frequency and amplitude.

\end{abstract}

\section{Introduction}
\setcounter{footnote}{0}
\label{intro}
The McCook-Sion white dwarf catalog contains more than 150 white dwarf stars with molecular or atomic carbon lines in their spectra (DQ white dwarfs; \citealt{mcc99} and online updates).  These lines often dominate DQ spectra, yet they are typically modeled by atmospheres containing mostly helium \citep{duf05}.  Recently, however, calculations by \citet{duf07} revealed that a small number of the hottest DQ white dwarf stars have spectra that are fit best by carbon-dominated model atmospheres.  These hot DQ white dwarf stars have shown themselves to be intriguing in other regards as well.  It is not, for instance, clear where they came from, though recent theoretical work suggests a scenario in which they descended from the hydrogen-deficient PG 1159 stars \citep{alt09, cor09}.

Adding to their intrigue and also promising to shed light on their origin and nature, recent observations have uncovered variables among the hot DQ white dwarf stars.  Compelled by the idea that carbon-atmosphere white dwarf stars might be pulsationally unstable, \citet{mon08} observed six hot DQ white dwarfs to look for photometric variability and discovered significant periodic luminosity variations in \dqtx.  \citet{bar08} then found variability in two more hot DQ white dwarfs (\dqone\ and \dqtwo).  \citet{gre09} and \citet{duf09} present follow-up observations of these three stars.

We here introduce a fourth hot DQ variable, SDSS J133710.19--002643.6 (hereafter \dqzero), a star first identified as a DQ white dwarf by \citet{lie03} and one of the nine hot DQ stars discussed by \citet{duf08}\footnote{Listed there with erroneous positive declination: SDSS J133710.19+002643.7}.   \dqzero\ is the fastest, smallest-amplitude variable of the four.  We detect two harmonically-related frequencies:  a 339 s fundamental and a 169 s first harmonic of comparable amplitude ($\sim$ 0.3\%).  Their phase relationship is like those of \dqtx\ and \dqone\ and leads to both a deep primary minimum and a secondary minimum at the location of the fundamental maximum.

\section{Observations}
\label{phot}
We observed \dqzero\ (g'=18.7, u'-g'= -0.46) using two different instruments on the 4.1-m SOAR telescope on Cerro Pachon in Chile.  For the first two sets of observations, we used the Goodman Spectrograph \citep{cle04}, an imaging spectrograph mounted at one of the SOAR Nasmyth ports.  The spectrograph's collimator and camera optics produce an image on a 4k $\times$ 4k Fairchild 486 back-illuminated CCD.  The plate scale at the detector is 0.15 arcsec pixel$^{-1}$.

\begin{table*}
\caption{\textsc{Observation Log}}
\centering
\begin{tabular}{ccccccccc}
\hline
\hline
Date & Start Time & T$_{exp}$ & T$_{cycle}$ & Length & Airmass & Instrument & Filter & Comparison Stars\\
(UTC) & (UTC) & (s) & (s) & (s) & & & & \\
\hline
2008 Jul 27 & 00:00:59.4 & 45 & 49.5 & 5939  & 1.34--2.01 & Goodman & none    & C1\tablenotemark{a},C2\tablenotemark{b}\\
2009 Apr 20 & 05:24:24.9 & 40 & 44.0 & 12878 & 1.19--3.11 & Goodman & S8612   & C1,C2\\
2009 Jun 27 & 23:57:00.3 & 38 & 39.9 & 13390 & 1.16--2.09 & SOI     & SDSS g' & C1,C2,C3\tablenotemark{c}\\
2009 Jul 23 & 23:54:00.4 & 25 & 27.9 & 9947  & 1.28--2.92 & SOI     & SDSS g' & C1,C2,C3\\
\hline
%\multicolumn{9}{ccccccccc}
$^{a}$SDSS J13:37:13.17--00:28:33.3 &  g'=16.9 & u'-g'=1.5 & & & & & &\\
$^{b}$SDSS J13:37:12.50--00:26:11.7 &g'=17.0 & u'-g'=1.2 & & & & & &\\
$^{c}$SDSS J13:37:18.44--00:25:58.3 &g'=17.1 & u'-g'=1.9 & & & & & & \\
\end{tabular}
\centering

\label{table:obs}
\end{table*}

During the 2009 June and July observations, the spectrograph was not available, so we used the SOAR Optical Imager (SOI; \citealt{wal03, sch04}).  The imager is mounted at a bent-Cassegrain port of the telescope and has a mosaic of two e2v 2k $\times$ 4k CCDs read out through four amplifiers.  The optics produce a plate scale at the detectors of .0767 arcsec pixel$^{-1}$.

We first obtained usable data on \dqzero\ on 2008 July 27 during an engineering night for the spectrograph.  We observed a field containing the target with the Goodman Spectrograph in imaging mode with the CCD readout binned 2 $\times$ 2, yielding 0.3 arcsec pixels.  We collected 1.6 hrs of usable unfiltered photometry during which the average seeing was 2.8 arcsec and became increasingly unstable. To decrease the read-out time, we obtained only a 300 $\times$ 650 pixel region of interest (ROI).  On 2009 April 20, we observed \dqzero\ for 3.6 hrs with the same setup as before but through a broadband blue S8612 filter and with a 600 $\times$ 280 ROI.  The average seeing was 1.3 arcsec, and the second half of the data, as the Moon was rising, shows significant periodic variations in the sky brightness with a period of around 720 s, as might result from passing clouds with periodic structure.

The next two observing nights presented us with more stable atmospheric conditions.  On 2009 June 27, we obtained 3.7 hrs of data through a Sloan g' filter on SOI.  The whole chip was read out and binned 6 $\times$ 6 to yield 0.46 arcsec pixels, which oversampled the poor average seeing of 2.5 arcsec.  We gathered 2.8 hrs of data with SOI on 2009 July 23 with a $\sim$ 5 min gap due to a guiding problem.  The 4 $\times$ 4 binning resulted in a plate scale of 0.31 arcsec pixel$^{-1}$, and the average seeing was 1.1 arcsec.  Table \ref{table:obs} summarizes these observations.

\section{Reduction and Analysis}
\label{reduct}
We used IRAF to subtract the bias level and apply flat field corrections to the Goodman data.  The unfiltered dome flats used to correct the first data set clearly do not correct obvious fringing structure present in the observation frames.  This fringing introduces errors into the photometry for that night, but its consequences are abated somewhat by the stability of the fringing pattern during the observations and the small, incremental drift of only about one pixel during the observing run.

\begin{figure*}
\epsscale{1}
\plotone{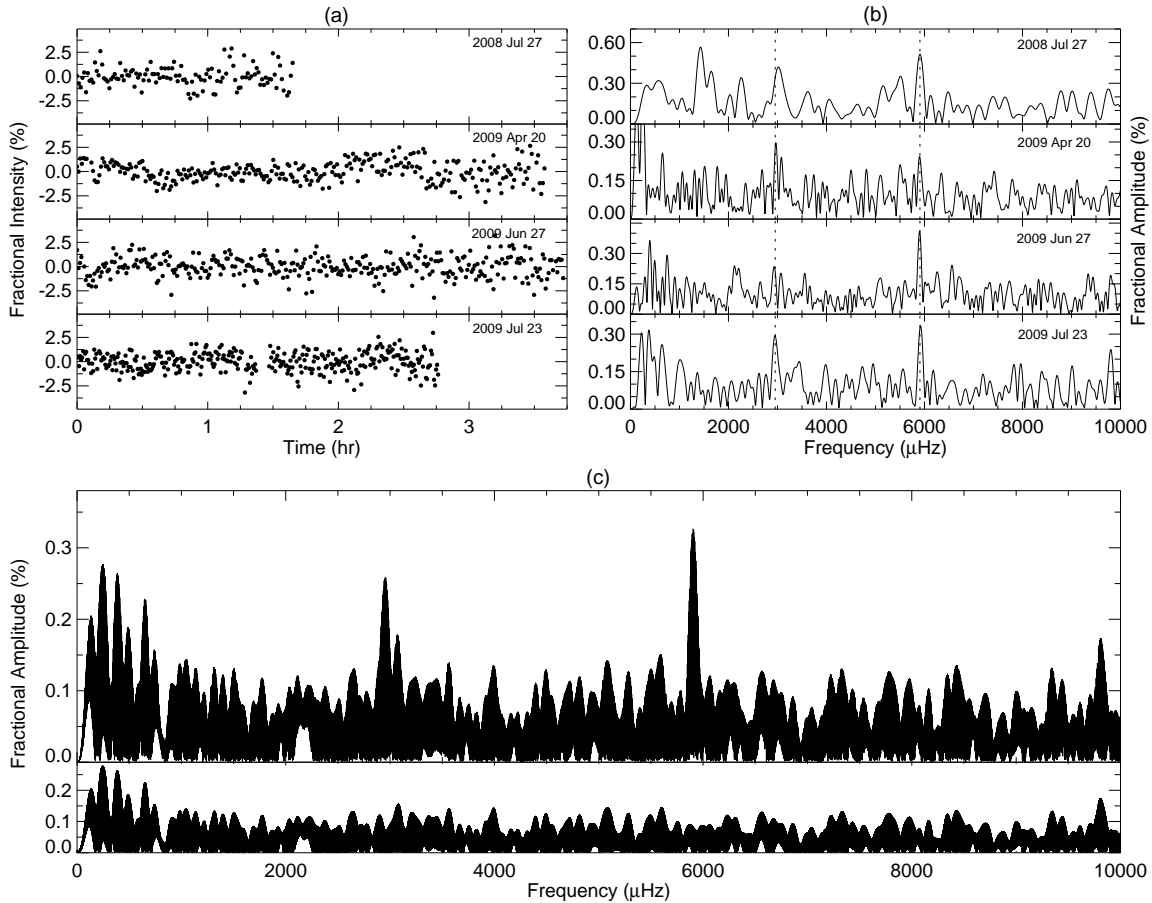}
\caption{(a) Differential light curves for each night and (b) amplitude spectra resulting from their Fourier transforms.  Dotted lines indicate the positions of the two detected frequencies.  (c)  Top:  Amplitude spectrum (produced with Period04) from the three nights of 2009 data combined. Significant peaks are present near the harmonically-related frequencies of 2950 $\mu$Hz (0.26\%) and 5900 $\mu$Hz (0.33\%).  Bottom:  Amplitude spectrum of the residuals that result from subtracting the best-fit model (with f$_{1}$ $\approx$ 2950.856 $\mu$Hz and f$_{2}$ = 2f$_{1}$) from the combined light curve.  No peaks above 700 $\mu$Hz are \textgreater 4 times the mean noise level of 0.055\%.}
\label{fig:lcs}
\end{figure*}

The data from SOI are written in mosaic format since each image consists of data from four different amplifiers with distinct bias levels and gains.  We used IDL to trim the overscan regions from each of the four data sections and write each exposure into one single-format FITS file (ignoring the gap between the two CCDs).  Once this was done for all the object, bias, and flat frames, we reduced the data using IRAF.  The resultant images display residual differences in mean bias level among the regions read out by the four amplifiers, but since we only care about differential photometry, it seemed superfluous to employ more manipulative means to remove this artifact.

We analyzed the data by first using a differential photometry program we wrote in IDL that uses the function APER, which is based on DAOPHOT \citep{ste87}.  We computed signal-to-noise (S/N) estimates with different-sized apertures and used the aperture that maximizes the S/N in the light curve \citep{how89} to perform aperture photometry using the program CCD\_HSP, an IRAF routine written by Antonio Kanaan that also uses DAOPHOT.

The program WQED (v. 2.0, \citealt{tho09}) was used to correct the times from UTC to the barycentric Julian ephemeris date and to carry out the following steps.  To mollify the effects of atmospheric transparency variations and extinction, we divided the light curve of \dqzero\ by the average of the light curves of comparison stars (Table \ref{table:obs}), which we checked against each other for signs of variability.  We fit a parabola to this divided light curve to approximate residual atmospheric extinction effects.  Dividing by this fit and subtracting one yields the light curves in terms of fractional variation about the mean presented in Figure \ref{fig:lcs}a.

Though no signal is apparent to the eye in the photometry, the amplitude spectra (Fig. \ref{fig:lcs}b) produced from discrete Fourier transforms of each of the light curves reveal noticeable signals on all four nights near 5900 $\mu$Hz (169 s).  A peak near the harmonically-related frequency 2950 $\mu$Hz (339 s) does not always stand out, but the combined 2009 amplitude spectrum shows both to be obviously above the noise (Fig. \ref{fig:lcs}c).

We assess the significance of these peaks from the combined 2009 data using the Lomb-Scargle normalized periodogram \citep{lom76, sca82} computed by IDL's LNP\_TEST (a routine based on fasper from \citealt{pre92}).  The largest peak in the power spectrum produced from the three nights of 2009 data combined is near 5900 $\mu$Hz and has a power of $\sim$ 26.  If we expect from the 2008 data a peak near $\sim$ 5900 $\mu$Hz and consider just this frequency, then the probability calculation is straightforward.  When the periodogram is normalized by the sample variance, the distribution of powers is described by the regularized incomplete beta function \citep{sch98} from which we find that the probability that a peak as large as the one near $\sim$ 5900 $\mu$Hz would occur there by chance is $\sim$ 3$\times$10$^{-12}$.

If, on the other hand, we want to know the false alarm probability, i.e., the odds of a peak so large occurring by chance at some frequency in the range considered (0--11370 $\mu$Hz, the Nyquist frequency on the April night), then we need to know the number of independent frequencies that serve to increase the probabilistic resources and thus increase the odds of finding a large peak due to noise.  We follow the method laid out in section 3.4.1 and Appendix B of \citet{cum99} (see also \citealt{hor86}) and perform 10$^5$ bootstrap Monte Carlo trials.  For each trial we compute the power spectrum of a light curve constructed with the same observation times as the original but with the flux values for each time drawn randomly (with replacement) from the original flux values.  A fit to the high probability end of the resulting distribution of maximum powers indicates that the number of independent frequencies is $\sim$ 9 times the number of data points (9$\times$961).  This results in a false alarm probability of $\sim$ 3$\times$10$^{-8}$, which is a factor of 10 smaller than an extrapolation of the Monte Carlo results.

Similarly, the probability that a peak as large as the one near 2950 $\mu$Hz ($\sim$ 16) will occur there by chance is $\sim$ 7$\times$10$^{-8}$.  The probability of a chance occurrence of a peak at least this high somewhere in the frequency range is $\sim$ 0.07\% according to the analytic calculation, or $\sim$ 0.15\% according to the Monte Carlo results.

To characterize these variations, we model them as a sum of sine waves and determine the best-fit amplitude, frequency, and phase by non-linear least-squares fits to the data, which we performed using both Period04 \citep{len05} and MPFIT \citep{mar09}.  The largest peak in the 2008 July amplitude spectrum is at approximately half the 2950 $\mu$Hz frequency but is not significantly present on the subsequent nights, so we include this extra, low-frequency component in the fit for that night only and note that fitting without it does not yield a significant difference. We list the best-fit parameters and their formal errors in Table \ref{table:fit1}.

\begin{table}
\caption{\textsc{Best-fit Parameters}}
\centering
\begin{tabular}{lccc}
\hline
\hline
Period & Frequency & Amplitude \\
(s) & ($\mu$Hz) & (\%) \\
\hline
\multicolumn{3}{l}{2008 Jul 27:}\\
\hspace{5pt}  700.7  $\pm$ 9.2  & 1427 $\pm$ 19 & 0.55 $\pm$ 0.11\\
\hspace{5pt}  331.7  $\pm$ 2.8  & 3015 $\pm$ 26 & 0.40 $\pm$ 0.11\\
\hspace{5pt}  169.30 $\pm$ 0.59 & 5907 $\pm$ 21 & 0.50 $\pm$ 0.11\\
2009 Apr 20: & & & \\
\hspace{5pt}  337.0  $\pm$ 1.4  & 2967 $\pm$ 12 & 0.293 $\pm$ 0.080\\
\hspace{5pt}  169.49 $\pm$ 0.41 & 5900 $\pm$ 14 & 0.243 $\pm$ 0.081\\
2009 Jun 27: & & & \\
\hspace{5pt}  340.9  $\pm$ 1.6  & 2933   $\pm$ 14  & 0.232 $\pm$ 0.077\\
\hspace{5pt}  169.47 $\pm$ 0.22 & 5900.7 $\pm$ 7.8 & 0.408 $\pm$ 0.077\\
2009 Jul 23: & & & \\
\hspace{5pt}  338.9  $\pm$ 1.5  & 2950 $\pm$ 13 & 0.288 $\pm$ 0.070\\
\hspace{5pt}  168.94 $\pm$ 0.33 & 5919 $\pm$ 12 & 0.329 $\pm$ 0.070\\
\hline
\end{tabular}
\label{table:fit1}
\end{table}

Both \dqtx\ and \dqone\ have non-sinusoidal light curves because of the presence of a fundamental and first harmonic.  In both of those stars, the harmonic minima coincide with the maxima and minima of the fundamental.  This results in a pulse shape with a deepened minimum at the location of the fundamental minimum; relatively large harmonic amplitudes (as in \dqone) produce secondary minima at the location of the fundamental maxima; and smaller harmonic amplitudes flatten the fundamental maxima (as in \dqtx).  We investigate the phase relationship between the two modes in \dqzero\ under the assumption that they are harmonically related (if they are not, the shape of the light curve does not repeat at the fundamental frequency).  To show that this assumption is consistent with the data, we compute a weighted average of the frequencies on the four nights using the inverse variances as weights \citep{tay97}.  This gives f$_{1}$ = 2957 $\pm$ 7 $\mu$Hz and f$_{2}$ = 5906 $\pm$ 6$\mu$Hz, which is consistent with f$_{2}$ = 2f$_{1}$.

We refit the data for each night applying this frequency constraint.  A weighted average of these results gives f$_{1}$ = 2953.6 $\pm$ 2.7 $\mu$Hz (338.57 $\pm$ 0.30 s); the aliasing in the combined 2009 amplitude spectrum prevents us from confidently determining a more accurate frequency.  Consistent with the data, we assume the frequencies are the same on each night and again refit the light curves with the frequencies fixed to look for changes in amplitude and relative phase.  We multiply the best-fit amplitudes and their errors by $\pi$T$_{exp}$f/sin($\pi$T$_{exp}$f) to correct for the effect of a finite exposure time, T$_{exp}$ \citep{bal99}.  Table \ref{table:fit2} lists these results.  We report phase difference as the number of seconds between the minimum of the harmonic and the minimum/maximum of the fundamental and use negative values to indicate the harmonic minimum is shifted left of the fundamental minimum/maximum.  The one-sigma errors reported for the phase differences come from bootstrap Monte Carlo simulations, and in each case the value falls between the sum of the errors for the individual phases and the quadrature sum of those errors.

\begin{table}
\caption{\textsc{Best-fit Parameters with f$_{1}$ = 2953.6 $\mu$Hz \& f$_{2}$ = 2f$_{1}$}}
\scriptsize
\begin{tabular}{cccc}
\hline
\hline
Date & Amplitude\tablenotemark{a} of f$_{1}$ & Amplitude\tablenotemark{a} of f$_{2}$ & Phase Difference\\
(UTC) & (\%) & (\%) & (s) \\
\hline
2008 Jul 27 & 0.35 $\pm$ 0.13  & 0.58 $\pm$ 0.14 & +18 $\pm$ 24\\
2009 Apr 20 & 0.28 $\pm$ 0.08  & 0.26 $\pm$ 0.09 & -12 $\pm$ 21\\
2009 Jun 27 & 0.22 $\pm$ 0.08  & 0.44 $\pm$ 0.08 & -47 $\pm$ 23\\
2009 Jul 23& 0.29 $\pm$ 0.07  & 0.33 $\pm$ 0.07 & -1  $\pm$ 16\\
\hline

\end{tabular}
\footnotetext{The amplitudes and their errors have been multiplied by the factor given in the text to correct for finite exposure times.}
\label{table:fit2}
\end{table}

\begin{figure}
\epsscale{1.11}
\plotone{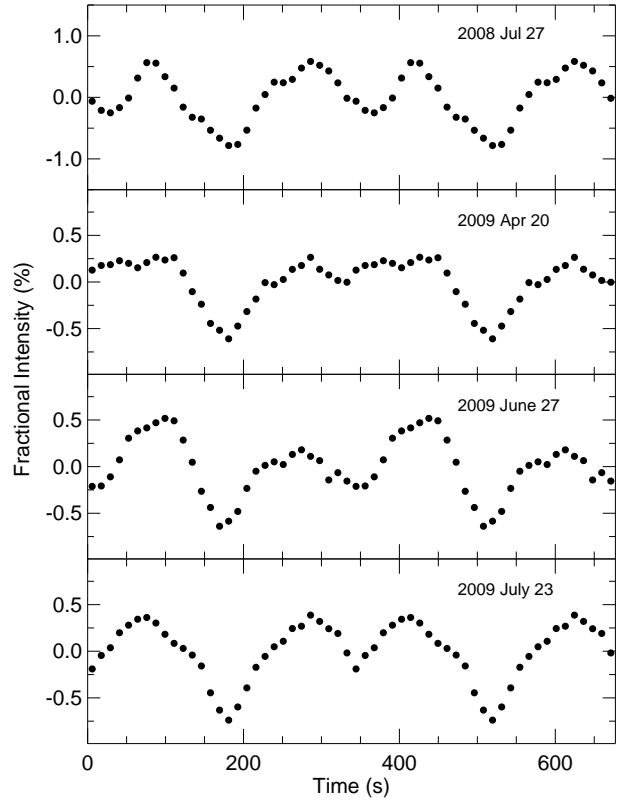}
\caption{Light curves for each of the four nights folded at the fundamental period.  The 1-$\sigma$ errors on the points range from $\sim$ 0.25\% to $\sim$ 0.50\%, but the scatter in the data does not represent this level of noise since our folding algorithm has the effect of smoothing the data.  The data are duplicated across two periods.}
\label{fig:flcs}
\end{figure}

By folding the light curves at the period of the fundamental, we get a picture of these quantitative results (Fig. \ref{fig:flcs}).  The pulse shape is like those of \dqtx\ and \dqone.  There is some suggestion that the phase relationship between the fundamental and harmonic might not be exactly zero, but the errors in phase are large making this hard to determine.  Similarly, there is no statistically significant change in amplitude or phase difference among the nights.

\section{Discussion}
\label{Disc}

The discovery of a fourth variable among the hot DQ white dwarf stars means the variable fraction of the known hot DQs is at least 36\%.  This is a high percentage compared with the hydrogen- and helium-atmosphere white dwarf variables.  Thus, as we get to know the hot DQ stars better, we find them disinclined to be photometrically constant.  Further, there is no obvious observational characteristic that serves as a predictor of whether a given hot DQ is variable.  Of the four reported variables, \dqtx\ has a spectral feature identified with helium \citep{duf08mag}; \dqzero\ has a spectral feature identified with hydrogen \citep{duf08}; two (\dqtx\ and \dqone) have broadened spectral lines \citep{duf08} possibly resulting from a magnetic field \citep{duf08mag}; the other two exhibit no such spectral distortions.  Thus, hydrogen, helium, and magnetic fields do not discourage variability, and none of them appears necessary to encourage it.  Indeed, so far, the best predictor of a hot DQ star's variability is that it is a hot DQ star.

Not only do we not know the reason one hot DQ varies and another doesn't, we do not know why their pulse shapes differ.  Based on suggestions of \citet{gre09}, \citet{duf09} predict a connection between magnetic field and pulse shape.  \dqtx\ and \dqone\ both have relatively large first harmonics whose minima coincide with the maxima and minima of the fundamental.  \dqtwo\ has no apparent harmonic.  \citet{duf09} and \citet{gre09} suggest that a magnetic field might account for this difference, but \dqzero\ calls this into question.  It has a pulse shape like that of \dqtx\ and \dqone, but unlike them its high S/N spectrum \citep{duf09spec} shows no signs of a magnetic field.

Other connections between the variables' spectroscopic properties and their variable properties are also not forthcoming.  The preliminary temperature fits of \citet{duf08} indicate that \dqzero\ is the hottest of the four variables; it is also the fastest and smallest in amplitude.  However, no obvious temperature-period trend emerges when considering the other three.  Such a relationship is observed and predicted in the ZZ Ceti pulsators \citep{muk06}.  Similarly, there is no straight-forward correlation between period and amplitude, a relationship also present in the ZZ Cetis \citep{muk06}.

We do know that all four variables have remained relatively stable in frequency and amplitude over the course of months to a year.  The variations observed in the present data set---the possible existence of a low-frequency peak in the 2008 July amplitude spectrum and the differences in best-fit phase difference and amplitude---are not conclusive.

Resolving these questions will require more observations.  High S/N photometry of the other hot DQ stars is required before any theoretical study can address which stars are variable and which are not.  \dqzero\ was observed in the original study of \citet{mon08} but not found to be variable on account of its small ($\sim$ 0.3\%) amplitude.  A mean amplitude spectrum noise level of even 0.1\% would not permit a convincing detection of such small-amplitude variability.  We also note that the 2009 June data show the harmonic at 4.1 times the mean noise in the amplitude spectrum while the fundamental is roughly half its size; thus, it seems possible that small-amplitude variable hot DQs with large harmonics could have fundamentals hidden in the noise.

In addition to high S/N time-series photometry, we need high S/N time-series spectroscopy.  With higher S/N spectroscopy, we might study line profile variations, which are a diagnostic of non-radial g-mode pulsations (e.g., \citealt{van00}), the leading mechanism for explaining hot DQ variability \citep{fon08, cor09}.  As with the ZZ Ceti pulsators, multi-color photometry will also be an important tool for demonstrating that the variability arises primarily from temperature variations \citep{rob82} and for determining the spherical harmonics of pulsational modes.  Unless more than one mode can be detected in each star, the prospects for seismology seem dim.  However, that was also the case for ZZ Ceti stars shortly after their discovery, and it is likely that aggressive campaigns to detect and observe more hot DQ variables will yield fruitful avenues of exploration that are richer and more interesting than we could have guessed.

\acknowledgments
We acknowledge the support of the National Science Foundation under award AST-0707381 and are grateful to the Abraham Goodman family for providing the financial support that made the spectrograph possible.  We thank the Delaware Asteroseismic Research Center for providing the S8612 filter used in these observations.  We also recognize the observational support provided by the SOAR operators Alberto Pasten, Patricio Ugarte, Sergio Pizarro, and Daniel Maturana.  BNB acknowledges support from a GAANN fellowship through grant number P200A090135 from the Dept. of Education.  BHD thanks GCC and the Melchior family for their role in this work.  Finally, we thank the referee for an enjoyable and helpful review process.
\\

\clearpage

\clearpage

\clearpage

\clearpage

\clearpage

\end{document}